\documentclass[a4paper,11pt]{article}
\usepackage{pos}
\usepackage{subcaption}
\usepackage{braket}
\newcommand{\phih}{\hat{\phi}}
\newcommand{\gket}[1]{|#1\rangle}
\newcommand{\gbra}[1]{\langle#1|}

\title{A Euclidean Monte-Carlo-informed route to ground-state preparation
for quantum simulation of scalar field theory}
\ShortTitle{Euclidean Monte-Carlo-informed ground-state preparation}

\author*[a,b,c]{Navya Gupta}
\author[b]{Christopher D. White}
\author[a,b,c]{Zohreh Davoudi}
\affiliation[a]{
Maryland Center for Fundamental Physics and Department of Physics, 
\\
University of Maryland, College Park, MD 20742, USA}
\affiliation[b]{Joint Center for Quantum Information and Computer Science,
\\
NIST/University of Maryland, College Park, \\
MD 20742 USA}
\affiliation[c]{The NSF Institute for Robust Quantum Simulation, University of Maryland, College Park, Maryland 20742, USA}

\emailAdd{navyag@umd.edu}
\abstract{Quantum simulators hold great promise for studying real-time (Minkowski) dynamics of quantum field theories. Nonetheless, preparing non-trivial initial states remains a major obstacle. Euclidean-time Monte-Carlo methods yield ground-state spectra and static correlation functions that can, in principle, guide state preparation. In this work, we exploit this classical information to bridge Euclidean and Minkowski descriptions for a (1+1)-dimensional interacting scalar field theory. We propose variational ansatz families (inspired by the stellar hierarchy for bosonic systems introduced in Ref.~\cite{chabaud2020stellar}) which achieve comparable ground-state energies, yet exhibit distinct correlations and local non-Gaussianity. By optimizing selected wavefunction moments with Monte-Carlo data, we obtain ansatzes that can be efficiently translated into quantum circuits. Our algorithmic cost analysis shows these circuits' gate complexity scales polynomially in system size. Our work paves the way for systematically leveraging classically-computed information to prepare initial states in quantum field theories of interest in nature.}

\FullConference{The 41st International Symposium on Lattice Field Theory (LATTICE2024)\\
 28 July - 3 August 2024\\
Liverpool, UK\\}

\begin{document}
\maketitle

\section{Introduction}

Real-time (Minkowski) dynamics of quantum field theories (QFTs) are notoriously difficult to address with standard Euclidean lattice methods due to a sign problem~\cite{troyer2005computational,cohen2015taming,goy2017sign}. Quantum simulators offer the exciting possibility of studying such dynamics~\cite{bauer2023quantum,halimeh2025quantum,bauer2023quantum2,di2024quantum,shaw2020quantum, ciavarella2021trailhead, kan2021lattice,lamm2019general,paulson2021simulating,davoudi2023general}. However, a major bottleneck  is efficiently preparing non-trivial initial states, such as ground or low-lying excited states~\cite{jordan2011quantum, jordan2012quantum,jordan2018bqp,kempe2006complexity,kitaev2002classical}. At the same time, path-integral Monte-Carlo (PIMC) techniques in Euclidean spacetime remain highly effective at extracting static observables in QFTs, including correlation functions and spectra~\cite{aoki2024flag,usqcd2019hot,davoudi2022report,kronfeld2022lattice,lin2022hadron} (when no sign or signal-to-noise problem hinders the computation). This raises the prospect of exploiting information about ground-state correlation functions obtained via Euclidean methods to guide or improve state preparation on quantum computers which operate in Minkowski time.

\begin{figure*}
    \centering
    \includegraphics[width=\textwidth]{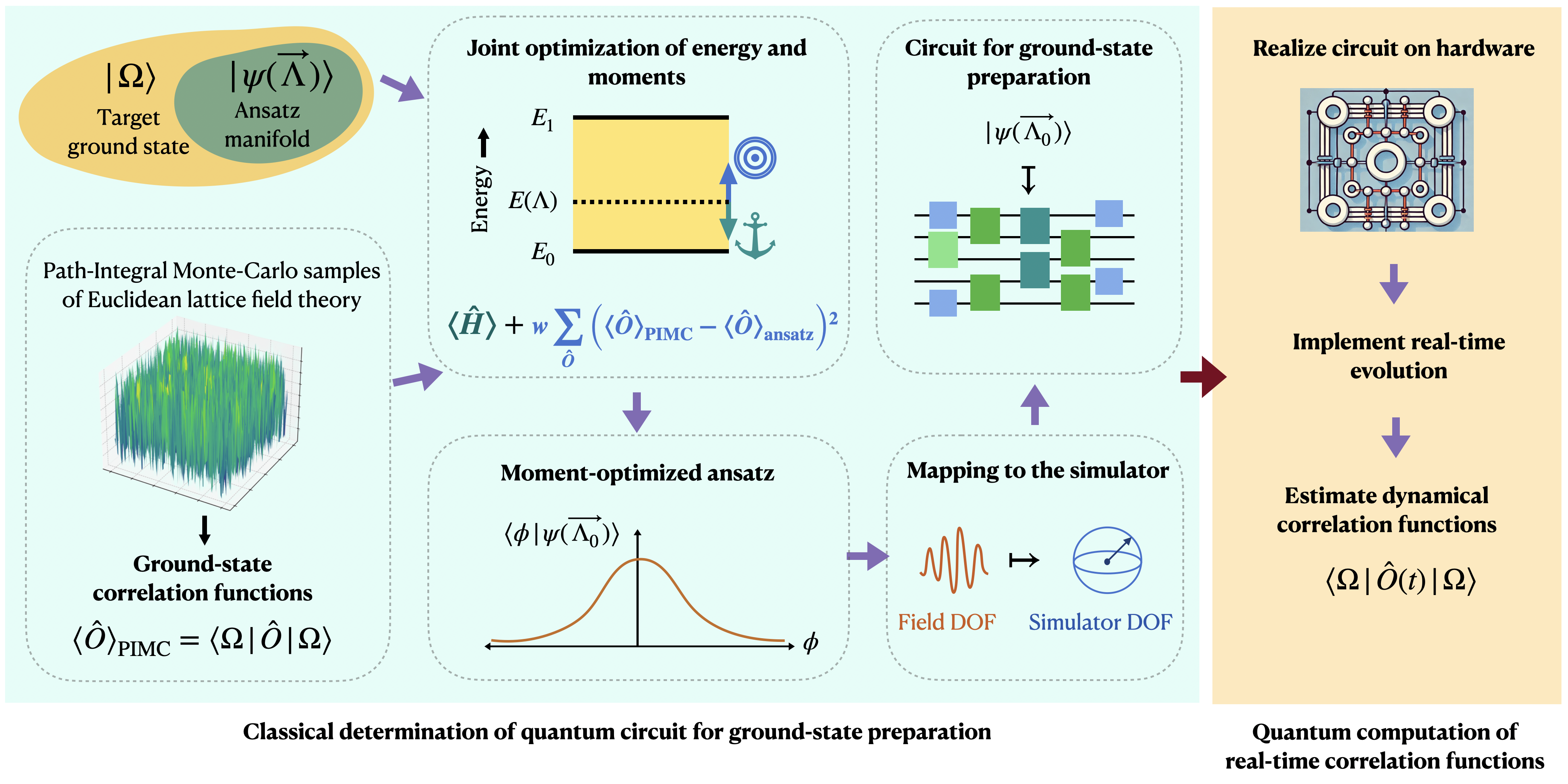}
    \caption{\textit{Schematic overview of classically informed ground-state preparation.} Using ground-state correlation functions in an interacting scalar field theory, sourced from Euclidean path-integral Monte Carlo, a joint optimization of the energy and moments of the ansatz wavefunction is performed. Based on a mapping between the field and simulator degrees of freedom, the optimized ansatz is translated into a quantum circuit using a classical algorithm. This classically determined circuit can thereafter be implemented on quantum hardware, and be used as the starting point for performing other tasks, such as simulating real-time dynamics and estimating dynamical correlation functions. The goal of the work presented is to demonstrate a classical determination of the quantum circuit for preparing the ground state.
    }
    \label{fig:pipeline}
\end{figure*}

In this work, we demonstrate a strategy to bridge these two pictures in the context of an interacting scalar field theory in (1+1) dimensions (see Fig.~\ref{fig:pipeline}). Specifically, we use PIMC data for ground-state correlation functions to inform the optimization of a variational family of wavefunction ansatzes. This variational ansatz is inspired by the stellar hierarchy for bosonic quantum systems~\cite{chabaud2020stellar,chabaud2022holomorphic}. The knowledge of ground-state correlation functions (i.e., moments) allows us to jointly optimize the energy and moments of this ansatz. Furthermore, we show that the ansatz admits an efficient mapping to a quantum circuit. This allows us to ``compile'' the optimized ansatz into a circuit whose gate complexity is polynomial in the lattice size and can be implemented on a qubit-based quantum computer. Thus, this method relies solely upon classical computing to determine the quantum circuit for implementing the ground state. Once this circuit is realized on quantum hardware, it can be used as an input for the quantum computation of dynamical correlation functions. 

In summary, the two main pillars of our work are: 1) A bosonic ansatz that is \emph{classically tractable} (i.e, can be optimized efficiently using classical resources), \emph{circuit translatable} (i.e., can be efficiently translated into a quantum circuit using classical resources), and \emph{circuit
efficient} (i.e, the resulting quantum circuit's complexity scales polynomially with system size); and 2) a Euclidean-Monte-Carlo-informed moment-optimization procedure for
optimizing this ansatz. The efficiency of our ansatz stems from the fact that it is directly expressed in terms of the degrees of freedom of the target field theory, making it straightforward to encode features of the theory (e.g., correlations, non-Gaussianity) in the ansatz.
Given that any ansatz will only express limited features of a non-trivial interacting ground state, we use Euclidean-Monte-Carlo-informed moment optimization to tune the ansatz to reproduce the most relevant features of the ground state. Such features depend on the subsequent quantum-simulation goals, e.g., studies
of certain excitations and their dynamics.

\section{The Model: (1+1)D $\phi^4$ theory}

The Hamiltonian for the continuum (1+1)D lattice \(\phi^4\) theory is given by
\begin{equation}
    \hat{H}_{\mathrm{cont}} \;=\; \int \! dx \,\Bigl[
    \tfrac12\,\hat{\pi}(x)^2
    \;+\; \tfrac12\,\bigl(\partial_x \hat{\phi}(x)\bigr)^2 
    \;+\; \tfrac12\,m_0^2\,\hat{\phi}(x)^2
    \;+\; \tfrac{\lambda_0}{4}\,\hat{\phi}(x)^4
    \Bigr],
\end{equation}
where $m_0^2$ and $\lambda_0$ are the bare mass and coupling, respectively, and the fields $\hat{\phi}$ and $\hat{\pi}$ satisfy the canonical bosonic commutation relation \([\hat{\phi}(x),\hat{\pi}(x')] = i\,\delta(x - x')\). We discretize space to a periodic lattice with spacing \(a\) and length \(L=Na\).  Defining the dimensionless fields \(\hat{\phi}_j \coloneq \hat{\phi}(j a)\) and \(\hat{\pi}_j \coloneq a\,\hat{\pi}(j a)\), the dimensionless lattice Hamiltonian reads
\begin{align}
    \hat{H} \;\coloneq\; a\,\hat{H}_{a,N}
    \;=\;\sum_{j=0}^{N-1} \Bigl[
    \tfrac12\,\hat{\pi}_j^2
    \;+\;\tfrac12\,\bigl(\hat{\phi}_{j+1} - \hat{\phi}_j\bigr)^2
    \;+\;\tfrac12\,m^2\,\hat{\phi}_j^2
    \;+\;\tfrac{\lambda}{4}\,\hat{\phi}_j^4
    \Bigr],
    \label{eq:hamiltonian}
\end{align}
where \(m^2 \coloneq m_0^2 a^2\) and \(\lambda \coloneq \lambda_0 a^2\). This Hamiltonian possesses ($\mathbb{Z}_2$) parity [i.e., $(\hat \phi, \hat \pi) \mapsto (-\hat \phi,-\hat \pi)$], time reversal, lattice inversion (i.e., spatial parity), and ($\mathbb{Z}_N$) lattice translation symmetries. For \(m^2<0\), the parity symmetry is spontaneously broken for a critical value of $\lambda$ in the thermodynamic limit \cite{loinaz1998monte}. The continuum limit is given by taking $N \rightarrow \infty$ followed by $(m^2,\lambda)\rightarrow (0,0)$. For this limit to converge, the couplings $(m^2,\lambda)$ must be tuned suitably with $a$~\cite{jordan2011quantum}.

 To keep our PIMC simple, we work with a modest lattice size \(N=10\). We choose \((m^2,\lambda)\) to ensure the lightest scalar mode fits well within the system volume (\(N \Delta E > 1\)) and remains below the UV cutoff (\(\Delta E < \pi\)), where $\Delta E$ denotes the spectral gap. Instead of determining the exact tuning of the couplings $(m^2,\lambda)$ with the lattice spacing $a$, we simply take a series of four values of \((m^2,\lambda)\) approaching \((0,0)\) along a straight line in the symmetric phase. 

PIMC is then used to estimate the two-point function \(\langle \hat{\phi}_0\,\hat{\phi}_j\rangle\), which characterizes non-local correlations in the ground state. Additionally, the moment ratio
\begin{equation}
    R_{2n} \;\coloneq\; \frac{\bigl\langle \hat{\phi}_j^{2n}\bigr\rangle}
    {(2n-1)!!\,\bigl\langle \hat{\phi}_j^2\bigr\rangle^n},
\end{equation}
a simple measure of the local non-Gaussianity of the ground state, is computed from local moments. 

\section{The $(R,Q)$ ansatz.} 

The ladder operators for the \(N\) bosonic modes are given by $\hat a_j \;\coloneq\; \frac{\hat{\phi}_j \,+\, i\,\hat{\pi}_j}{\sqrt{2}}$ and $\hat a_j^{\dagger} \;\coloneq\; \frac{\hat{\phi}_j \,-\, i\,\hat{\pi}_j}{\sqrt{2}}$;
they satisfy the commutation relations 
\([\hat{a}_j,\hat{a}_k^\dagger] \;=\;\delta_{j,k}\) and 
\([\hat{a}_j,\hat{a}_k] \;=\;0\). An $N$-mode bosonic state \(\gket{\psi}\) is said to have a \emph{finite stellar rank} \(R\) if it can be written as~\cite{chabaud2020stellar,chabaud2022holomorphic}
\begin{equation}\label{eq:multimode_stellar}
    \gket{\psi} \;=\; \hat{U}_{G}\,\gket{C}.
\end{equation}
Here, \(\hat{U}_{G}\) is a Gaussian unitary (i.e., is generated by a quadratic Hamiltonian in $\hat{a}_j$ and $\hat{a}_j^\dagger$). Furthermore, \(\gket{C} \coloneq C(\hat a_0^{\dagger},\cdots,\hat a_{N-1}^{\dagger})\,\gket{\mathbf{0}}\) is referred to as the \emph{core state}: it is generated from the Fock vacuum $\gket{\mathbf{0}}\coloneq  \gket{0,\cdots,N-1}$ by $C(\hat a_0^{\dagger},\cdots,\hat a_{N-1}^{\dagger})$---a polynomial in the boson creation operators with degree $R$. Bosonic states which do not admit the decomposition given in Eq.~\eqref{eq:multimode_stellar}---such as the $\phi^4$ theory's ground state---are said to have an infinite rank. Finite-rank states form a dense subset of the $N$-mode Hilbert space in trace distance. In other words, finite-rank states can get arbitrarily close to a given infinite-rank state in trace distance, provided their rank is large enough. This motivates us to search for ansatzes for representing the $\phi^4$ ground state within the space of finite-rank states.

The first constraint one may impose on rank-$R$ states is that they obey the symmetries of the $\phi^4$ Hamiltonian. We will simplify such symmetric rank-$R$ states even further to aid the classical optimization of the ansatz. In particular, we restrict the Gaussian unitary to be a product of single-mode squeezing operations, i.e.,
\begin{equation}
    \hat U_G=\bigotimes_{j=0}^{N-1} \hat{S}_j(r),
\end{equation}
where $\hat S_j(r)\coloneq e^{\frac{r}{2}\left((\hat a_j^{\dagger})^2-\hat a_j^2\right)}$ is the squeezing operator on mode $j$ with the real squeezing parameter $r$. As for the core state, it can be generated by a polynomial in the quadrature operators, i.e., $\gket{C}=C(\{\hat a_j^{\dagger}\})\gket{0} \equiv C^{\phi}(\{\hat \phi_j\})\gket{0}$. It is useful to express the state in terms of $C^{\phi}(\cdot)$ because we aim to compute expectation values of monomials involving $\hat \phi_j$ and $\hat \pi_j$. In addition to restricting the degree of $C^{\phi}(\cdot)$ to be $R$, we demand that it only consists of terms of the form $\hat\phi_j^{n_0}\hat \phi_{j+1}^{n_1}\cdots\hat\phi_{j+q}^{n_q}$ with $ n_0\geq 1$, $n_q\geq 1$, and $n_{j'}\geq 0$ for $j'\in \{1,\cdots,q-1\}$, where $q$ is less than or equal to some chosen truncation $Q\leq N/2$. 
On a periodic and inversion-symmetric lattice, the term $\hat\phi_j^{n_0}\hat \phi_{j+1}^{n_1}\cdots\hat\phi_{j+q}^{n_q}$ represents the shortest possible wrapping of the operators around the lattice.
Thus, $Q$ specifies the spread of the boson additions performed by the core-state polynomial. These two simplifications leave us with the \emph{$(R,Q)$ ansatz}
\begin{align}\label{eq:core_state_decomp}
    \gket{\psi}_{R,Q} \;&\coloneq\; 
    \Bigl[\!\bigotimes_{j=0}^{N-1} \hat{S}_j(r)\Bigr]\,
    C_{R,Q}\bigl(\hat{\phi}_0,\cdots,\hat{\phi}_{N-1}\bigr)\,\gket{0}\coloneq \Bigl[\!\bigotimes_{j=0}^{N-1} \hat{S}_j(r)\Bigr]\gket{C}_{R,Q},
\end{align}
with
\begin{align}
    C_{R,Q}\bigl(\hat{\phi}_0,\cdots,\hat{\phi}_{N-1}\bigr) &\coloneq \sum_{\substack{0\leq R'\leq R \\ R'\text{ is even} }}\sum_{0\leq q\leq Q}\sum_{\substack{n_0,n_1,\cdots,n_q \\ n_0,n_q\geq 1 \\ n_0+n_1+\cdots+n_q=R'}}c_{n_0,n_1,\cdots,n_q}\sum_{j=0}^{N-1}\hat\phi_j^{n_0}\phi_{j+1}^{n_1}\cdots\hat\phi_{j+q}^{n_q}.
\end{align}
The above ansatz manifestly possesses discrete translation invariance. Time-reversal invariance is ensured by choosing $r$ and polynomial coefficients $
c_{n_0,n_1,\cdots,n_q}$ to be real. Parity symmetry restricts the polynomial to only consist of even-degree terms. Finally, inversion symmetry implies that $c_{n_0,n_1,\cdots,n_q}=c_{N-n_0,N-n_1,\cdots,N-n_q}$. 

For comparison, we also consider the Gaussian Effective Potential (GEP) ansatz, which has been widely considered in the literature~\cite{stevenson1984gaussian,stevenson1985gaussian}. This is simply the ground state of the Hamiltonian in Eq.~\eqref{eq:hamiltonian} with $\lambda=0$, and with the mass $m=\mu\geq 0$ serving as the variational parameter.

\section{Euclidean-Monte-Carlo-informed moment optimization}

Suppose it is important for the ansatz $\gket{\psi(\vec{\Lambda})}$ to accurately reproduce the ground-state expectation values for some set of operators $\mathcal{T} = \{\hat O_i\}$, which we refer to as the ``target set.'' We propose to find the appropriate value of $\vec{\Lambda}$ by minimizing the objective function
\begin{align}\label{eq:mom_obj}
    \vec{\Lambda}_0 = \mathrm{argmin}_{\vec{\Lambda}} \ \left[\gbra{\psi(\vec{\Lambda})}\hat H \gket{\psi(\vec{\Lambda})} + \sum_{\hat O \in \mathcal{T}} \ w_{\hat O} \left(\gbra{\psi(\vec{\Lambda})}\hat O \gket{\psi(\vec{\Lambda})} - \gbra{\Omega} \hat O \gket{\Omega}\right)^2\right],
\end{align}
\noindent where the weights $w_{\hat O}\geq 0$ are determined by experimentation. We will refer to this procedure as \emph{Euclidean-Monte Carlo-informed moment optimization} since the ground-state moments $\gbra{\Omega}\hat O\gket{\Omega}$ will be sourced from PIMC. When $w_{\hat O}=0$, one recovers energy minimization. As the value of $w_{\hat O}$ is increased, the energy of the optimized ansatz grows. In a successful instance of moment optimization, the behavior of the ansatz target moments improves at sufficiently small values of the weights, such that the energy penalty paid is small with respect to the spectral gap. 

\begin{figure*}
    \includegraphics[width=\textwidth]{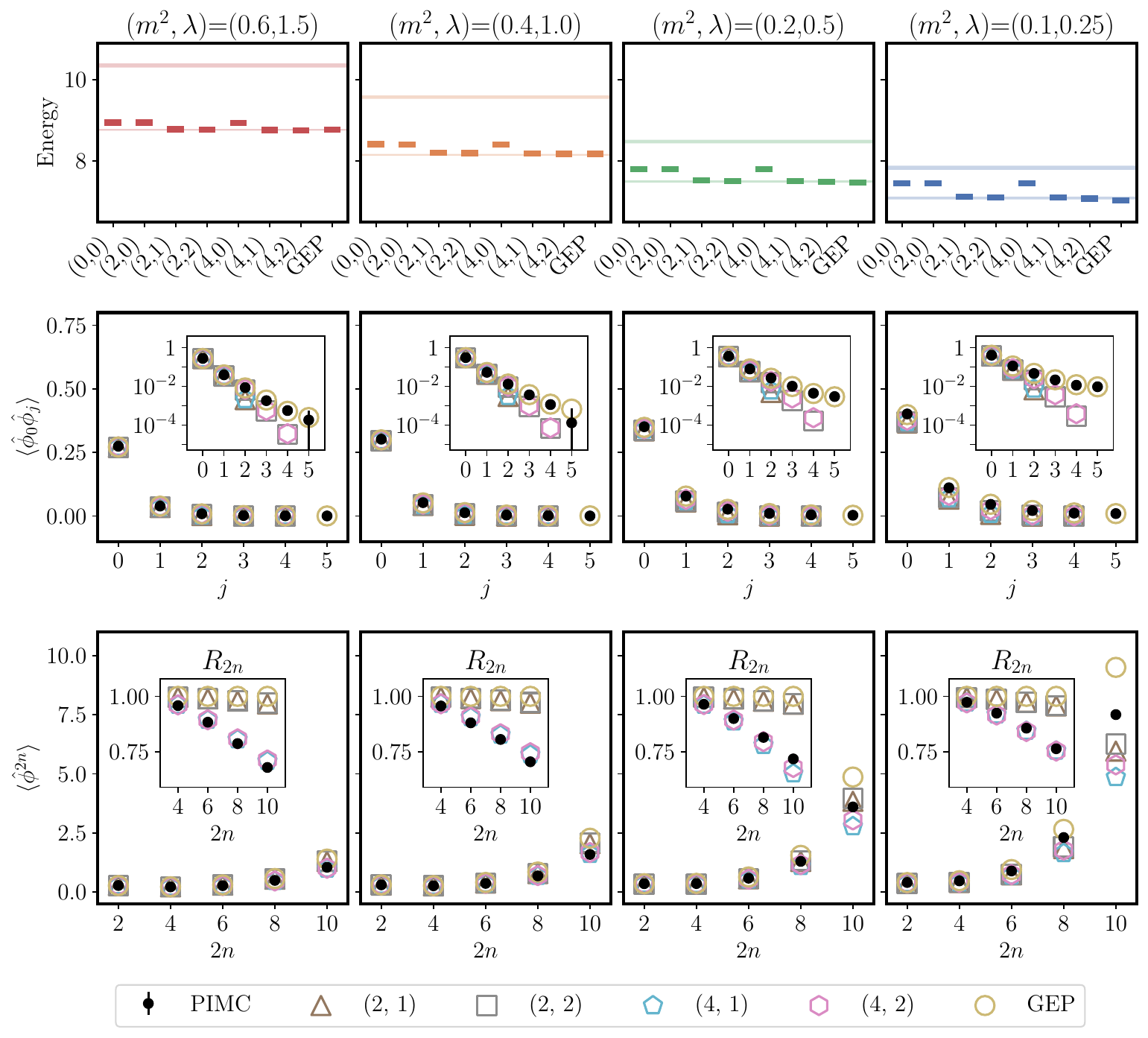}
    \caption{\textbf{Top:} Minimum energy of various ansatzes compared with the Monte-Carlo energy estimate for various values of $(m^2,\lambda)$. The two bands indicate the Monte-Carlo energy estimate for the ground state and first excited state. \textbf{Center:} Two-point function for various values of $(m^2,\lambda)$. \textbf{Bottom:} Local $\phi$-moments and moment ratio for the minimum energy $(R,Q)$ and GEP ansatzes for various values of $(m^2,\lambda)$. Black points with error bars in the center and bottom panels show the values from PIMC. While the GEP approximates the two-point function very well, it completely fails to capture the state's non-Gaussianity.}
    \label{fig:multimode_min_en}
\end{figure*}

As a baseline, we first consider energy minimization. The top row of Fig.~\ref{fig:multimode_min_en} shows that both the GEP and \((R,Q)\) ansatzes yield comparable values of energy close to the true ground-state energy. However, while the GEP reproduces two-point correlators well, it cannot reproduce non-Gaussian correlations by design. The \((R,Q)\) ansatzes, on the other hand, reproduce the moment ratio much more faithfully for $R= 4$, but their two-point correlation functions have a greater discrepancy with the ground-state value. All these trends can be seen in the bottom two rows of Fig.~\ref{fig:multimode_min_en}. To summarize, even though energy minimization leads to comparable values of energy for different ansatz families, the resulting minimum-energy ansatzes behave differently in terms of their non-local correlation functions and local non-Gaussianity. We will now explore how one can vary moments errors within a \emph{fixed} ansatz family using moment optimization.

\begin{figure}
    \centering
    \includegraphics[width=\columnwidth]{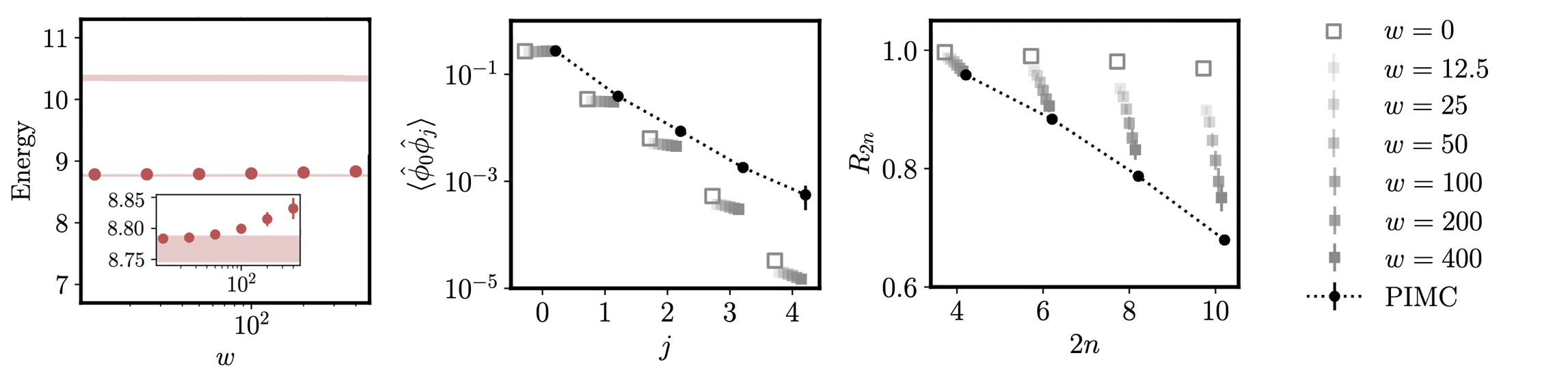}
    \caption{Optimization of the moment ratio for the $(R,Q)=(2,2)$ ansatz for $(m^2,\lambda)=(0.6,1.5)$. Moment optimization is performed for the target set $\mathcal{T}=\{\phih_j^6,\phih_j^8,\phih_j^{10}\}$ for various values of $w$. The behavior of the moment ratio improves continuously as $w$ is increased. The moment optimization results in a small increase in energy as a function of $w$. The behavior of the two-point function is also only slightly modified as compared to the minimum-energy case. The values associated with different weights and PIMC are slightly offset in the horizontal direction to
    improve visibility of the error bars.}
    \label{fig:moment_ratio_opt}
\end{figure}

\begin{figure}
    \centering
    \includegraphics[width=\columnwidth]{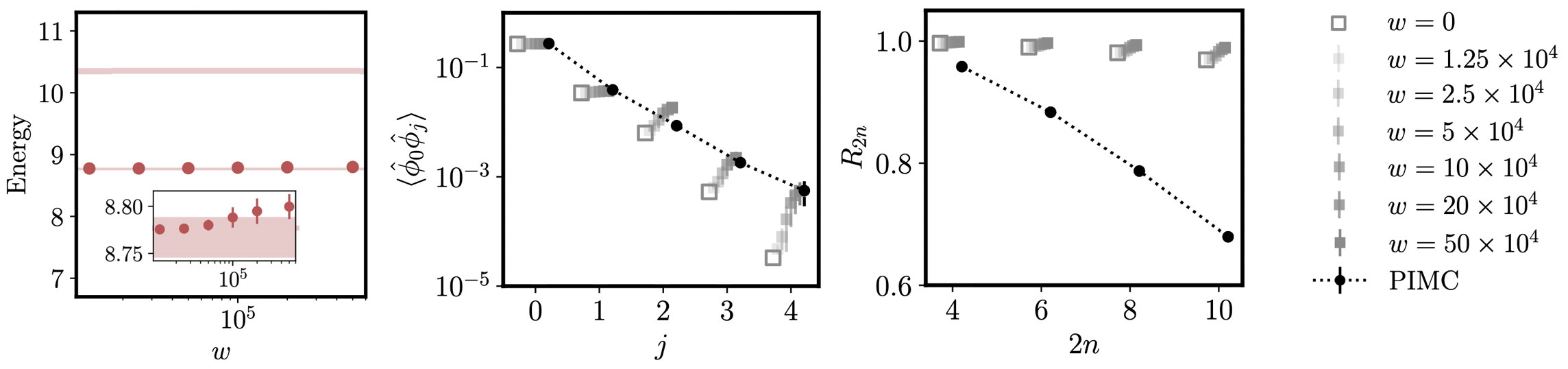}
    \caption{Optimization of the two-point function $\phih_0\phih_4$ for the $(R,Q)=(2,2)$ ansatz for $(m^2,\lambda)=(0.6,1.5)$. Moment optimization is performed for the target set $\mathcal{T}=\{\phih_0\phih_4\}$ for various values of $w$. The behavior of this two-point function improves continuously as $w$ is increased. The moment optimization results in a small increase in energy as a function of $w$. The behavior of the moment ratio is also only slightly modified as compared to the minimum-energy case. The values associated with different weights and PIMC are slightly offset in the horizontal direction to
    improve visibility of the error bars.}
    
    \label{fig:two_pt_opt}
\end{figure}

Using the objective function in Eq.~\eqref{eq:mom_obj}, we first optimize local higher-order moments (i.e., non-Gaussianities) using the target set $\mathcal{T}=\{\hat{\phi}_j^6,\hat{\phi}_j^8,\hat{\phi}_j^{10}\}$  for $(m^2,\lambda)=(0.6,1.5)$ and the $(R,Q)=(2,2)$ ansatz with a shared weight $w$ for all moments. As shown in Fig.~\ref{fig:moment_ratio_opt}, as $w$ increases, the moment ratios approach the PIMC value without a significant increase in energy and deviation in the two-point function values. Alternatively, as shown in Fig.~\ref{fig:two_pt_opt}, the target set $\mathcal{T}=\{\hat{\phi}_0\hat{\phi}_4\}$ can be used to improve the behavior of two-point function for the same values of $(m^2,\lambda)=(0.6,1.5)$ and the $(R,Q)=(2,2)$ ansatz. This shows that both the moment ratio and two-point correlation function at large separations can be separately improved with only a small cost in energy.

\section{Quantum-Circuit Representation\label{sec:circ}}

To map the optimized ansatz to a qubit-based quantum circuit, one must truncate the infinite-dimensional bosonic Hilbert space. In particular, we truncate at a maximum occupation \(\Lambda\) for each mode~\cite{somma2005quantum,macridin2018digital,klco2019digitization}, and encode the states \(\gket{0},\cdots,\gket{\Lambda}\) in a register of qubits using either a unary or binary map. Thus, $n_b(\Lambda)\coloneq\lceil \mathrm{log}(\Lambda+1)\rceil$ and $n_u(\Lambda)\coloneq \Lambda+1$ qubits are needed to represent each mode for the binary and unary maps, respectively.

\paragraph{Core-state preparation.}
The core state \(\gket{C}_{R,Q}\) in Eq.~\eqref{eq:core_state_decomp} is a superposition of $N|c|_{R,Q}$ Fock states, where the total occupation of each Fock state is no more than the rank $R$. Here, $|c|_{R,Q}$ is some combinatorial factor depending upon $R$ and $Q$. With a truncation \(\Lambda \geq R\), the core state can be prepared exactly. This procedure amounts to preparing a sparse superposition of the $N|c|_{R,Q}\ll2^{Nn_q}$ computational basis states that map to the Fock states created by the terms in the polynomial in Eq.~\eqref{eq:core_state_decomp}. The algorithm introduced in Ref.~\cite{gleinig2021efficient} prepares sparse states by determining a series of controlled ``merges'' that convert the desired superposition to a single computational basis state. Reversing these merges then yields the desired superposition. This strategy yields a quantum circuit with $O\left(N^2n_q(R)|c|_{R,Q}\right)$ CNOT gates in $O\left(N^3n_q(R)|c_{R,Q}|^2\mathrm{log}(N|c_{R,Q}|)\right)$ classical computation time. Once the ansatz form is chosen, only the rotation angles in these merges depend on the physical parameters \((m^2,\lambda)\).\\

\paragraph{Gaussian unitary.} We define the truncated single-mode squeezing operator $\hat S_j^{\Lambda}(r)$ as the  exponentiation of a squeezing Hamiltonian truncated at occupation number $\Lambda$. Since this Hamiltonian consists of non-commuting terms, one can employ a first-order Trotter-Suzuki product formula~\cite{trotter1959product,suzuki1976generalized} to approximate it. Thus, the implementation of the squeezing operator incurs errors both due to the Fock-space truncation and Trotterization. Once the value of the cutoff $\Lambda$ and the number of Trotter layers $K$ are fixed, the tensor product of these approximate-digitized squeezing operators can be implemented using $O(N\Lambda K)$ CNOT gates (assuming the unary mapping). For the binary mapping, the Singular Value Decomposition (SVD) algorithm introduced in Ref.~\cite{davoudi2023general} can be used to implement each Trotter layer with $O(N\Lambda K\mathrm{log}\Lambda)$ CNOT gates. 

\section{Conclusion}

In this work, we have demonstrated a method for determining quantum circuits that prepare states close to the ground state of the (1+1)D $\phi^4$ theory. We use ground-state correlation functions sourced from Euclidean path-integral Monte Carlo  to inform a moment-optimization procedure for pinning down ansatz parameters. Thereafter, the optimized ansatz is mapped to a quantum circuit. These circuits can be subsequently used as inputs for quantum simulation of real-time dynamics. The moment-optimization procedure uses the ground-state correlation functions to guide the optimization process to regions of the ansatz manifold which not only minimize energy discrepancy with the ground state, but also achieve accurate behavior of some target-moment sets. The extent to which various target-moment sets can be optimized depends upon the structure and expressiveness of the ansatz, together with the properties of the theory itself (such as its spectral gap).

The crucial component for translating Euclidean ground-state data to Minkowski wavefunctions is the finite stellar-rank ansatz, which we specialize to the $(R,Q)$ ansatz. The larger the values of $R$ and $Q$, the more expressive the ansatz. An important assumption of our work is that $R$ and $Q$ values are much smaller than the system size $N$. This makes ansatz optimization and quantum-circuit translation efficient. However, as the continuum limit is approached, we do expect that larger values of $R$ and $Q$ may be needed to achieve the same level of accuracy. Thus, it is important to examine the thermodynamic and continuum limits in more detail in the future. 

Optimizing the choice of weights in the moment-optimization objective function is another important area for further study. In our current work, these weights were selected through trial and error. However, higher-order moments in the target set can increase the tendency for excited-state contamination and are associated with higher statistical errors. Developing a principled approach to selecting moment-dependent weights could enhance the effectiveness of the optimization process. Moreover, since the ground-state correlation functions are sourced from Monte-Carlo simulations, the moment-optimized parameters inherently carry statistical errors. Understanding the implications of these errors for quantum simulations will be crucial. 

The long-term vision of this program is to extend the paradigm of Euclidean-Monte Carlo-informed ground-state preparation to theories involving fermions and gauge bosons. The hope is to incorporate the wealth of data about spectra and static correlation functions from lattice QCD towards preparing non-trivial initial states for quantum simulations of QCD.

\section{Acknowledgments}

N.~G. was supported by the U.S. National Science Foundation's Quantum Leap Challenge Institute (OMA-2120757). N.~G. and Z.~D. were supported, in part, by Maryland Center for Fundamental Physics, Department of Physics, and College of Computer, Mathematical, and Natural Sciences at the University of Maryland College Park. Z.~D. further acknowledges support by the U.S. Department of Energy (DOE), Office of Science, Early Career Award (DE-SC0020271), and the U.S. DOE, Office of Science, Office of Advanced Scientific Computing Research, Accelerated Research in Quantum Computing program: Fundamental Algorithmic Research toward Quantum Utility (FAR-Qu). C.D.W. gratefully acknowledges support from the U.S. Department of Energy (DOE), Office of Science, Office of Advanced Scientific Computing Research (ASCR) Quantum Computing Application Teams program, under fieldwork proposal number ERKJ347, 
as well as DOE Quantum Systems Accelerator program, DE-AC02-05CH11231, AFOSR MURI FA9550-22-1-0339, ARO grant W911NF-23-1-0242, ARO grant W911NF-23-1-0258, and NSF QLCI grant OMA-2120757.

\bibliographystyle{JHEP}
\bibliography{references.bib}

\providecommand{\href}[2]{#2}\begingroup\raggedright\begin{thebibliography}{10}

\bibitem{chabaud2020stellar}
U.~Chabaud, D.~Markham and F.~Grosshans, \emph{Stellar representation of non-gaussian quantum states}, {\emph{Physical Review Letters} {\bfseries 124} (2020) 063605}.

\bibitem{troyer2005computational}
M.~Troyer and U.-J.~Wiese, \emph{Computational complexity and fundamental limitations to fermionic quantum monte carlo simulations}, {\emph{Physical review letters} {\bfseries 94} (2005) 170201}.

\bibitem{cohen2015taming}
G.~Cohen, E.~Gull, D.R.~Reichman and A.J.~Millis, \emph{Taming the dynamical sign problem in real-time evolution of quantum many-body problems}, {\emph{Physical review letters} {\bfseries 115} (2015) 266802}.

\bibitem{goy2017sign}
V.~Goy, V.~Bornyakov, D.~Boyda, A.~Molochkov, A.~Nakamura, A.~Nikolaev et~al., \emph{Sign problem in finite density lattice qcd}, {\emph{Progress of Theoretical and Experimental Physics} {\bfseries 2017} (2017) 031D01}.

\bibitem{bauer2023quantum}
C.W.~Bauer, Z.~Davoudi, N.~Klco and M.J.~Savage, \emph{Quantum simulation of fundamental particles and forces}, {\emph{Nature Reviews Physics} (2023) 1}.

\bibitem{halimeh2025quantum}
J.C.~Halimeh, N.~Mueller, J.~Knolle, Z.~Papi{\'c} and Z.~Davoudi, \emph{Quantum simulation of out-of-equilibrium dynamics in gauge theories}, {\emph{arXiv preprint arXiv:2509.03586} (2025) }.

\bibitem{bauer2023quantum2}
C.W.~Bauer, Z.~Davoudi, A.B.~Balantekin, T.~Bhattacharya, M.~Carena, W.A.~de~Jong et~al., \emph{Quantum simulation for high-energy physics}, {\emph{PRX Quantum} {\bfseries 4} (2023) 027001}.

\bibitem{di2024quantum}
A.~Di~Meglio, K.~Jansen, I.~Tavernelli, C.~Alexandrou, S.~Arunachalam, C.W.~Bauer et~al., \emph{Quantum computing for high-energy physics: State of the art and challenges}, {\emph{Prx quantum} {\bfseries 5} (2024) 037001}.

\bibitem{shaw2020quantum}
A.F.~Shaw, P.~Lougovski, J.R.~Stryker and N.~Wiebe, \emph{Quantum algorithms for simulating the lattice schwinger model}, {\emph{Quantum} {\bfseries 4} (2020) 306}.

\bibitem{ciavarella2021trailhead}
A.~Ciavarella, N.~Klco and M.J.~Savage, \emph{{Trailhead for quantum simulation of SU(3) Yang-Mills lattice gauge theory in the local multiplet basis}}, {\emph{Physical Review D} {\bfseries 103} (2021) 094501}.

\bibitem{kan2021lattice}
A.~Kan and Y.~Nam, \emph{Lattice quantum chromodynamics and electrodynamics on a universal quantum computer}, {\emph{arXiv preprint arXiv:2107.12769} (2021) }.

\bibitem{lamm2019general}
H.~Lamm, S.~Lawrence, Y.~Yamauchi, N.~Collaboration et~al., \emph{General methods for digital quantum simulation of gauge theories}, {\emph{Physical Review D} {\bfseries 100} (2019) 034518}.

\bibitem{paulson2021simulating}
D.~Paulson, L.~Dellantonio, J.F.~Haase, A.~Celi, A.~Kan, A.~Jena et~al., \emph{Simulating 2d effects in lattice gauge theories on a quantum computer}, {\emph{PRX Quantum} {\bfseries 2} (2021) 030334}.

\bibitem{davoudi2023general}
Z.~Davoudi, A.F.~Shaw and J.R.~Stryker, \emph{General quantum algorithms for hamiltonian simulation with applications to a non-abelian lattice gauge theory}, {\emph{Quantum} {\bfseries 7} (2023) 1213}.

\bibitem{jordan2011quantum}
S.P.~Jordan, K.S.~Lee and J.~Preskill, \emph{Quantum computation of scattering in scalar quantum field theories}, {\emph{arXiv preprint arXiv:1112.4833} (2011) }.

\bibitem{jordan2012quantum}
S.P.~Jordan, K.S.~Lee and J.~Preskill, \emph{Quantum algorithms for quantum field theories}, {\emph{Science} {\bfseries 336} (2012) 1130}.

\bibitem{jordan2018bqp}
S.P.~Jordan, H.~Krovi, K.S.~Lee and J.~Preskill, \emph{Bqp-completeness of scattering in scalar quantum field theory}, {\emph{Quantum} {\bfseries 2} (2018) 44}.

\bibitem{kempe2006complexity}
J.~Kempe, A.~Kitaev and O.~Regev, \emph{The complexity of the local hamiltonian problem}, {\emph{Siam journal on computing} {\bfseries 35} (2006) 1070}.

\bibitem{kitaev2002classical}
A.Y.~Kitaev, A.~Shen and M.N.~Vyalyi, \emph{Classical and quantum computation}, no.~47, American Mathematical Soc. (2002).

\bibitem{aoki2024flag}
Y.~Aoki, T.~Blum, S.~Collins, L.~Del~Debbio, M.~Della~Morte, P.~Dimopoulos et~al., \emph{Flag review 2024}, {\emph{arXiv preprint arXiv:2411.04268} (2024) }.

\bibitem{usqcd2019hot}
U.~Collaboration, A.~Bazavov, F.~Karsch, S.~Mukherjee and P.~Petreczky, \emph{{Hot-dense lattice QCD}}, {\emph{The European Physical Journal A} {\bfseries 55} (2019) 1}.

\bibitem{davoudi2022report}
Z.~Davoudi, E.T.~Neil, C.W.~Bauer, T.~Bhattacharya, T.~Blum, P.~Boyle et~al., \emph{Report of the snowmass 2021 topical group on lattice gauge theory}, {\emph{arXiv preprint arXiv:2209.10758} (2022) }.

\bibitem{kronfeld2022lattice}
A.S.~Kronfeld, T.~Bhattacharya, T.~Blum, N.H.~Christ, C.~DeTar, W.~Detmold et~al., \emph{{Lattice QCD and particle physics}}, {\emph{arXiv preprint arXiv:2207.07641} (2022) }.

\bibitem{lin2022hadron}
H.-W.~Lin, \emph{Hadron spectroscopy and structure from lattice qcd}, {\emph{Few-Body Systems} {\bfseries 63} (2022) 65}.

\bibitem{chabaud2022holomorphic}
U.~Chabaud and S.~Mehraban, \emph{Holomorphic representation of quantum computations}, {\emph{Quantum} {\bfseries 6} (2022) 831}.

\bibitem{loinaz1998monte}
W.~Loinaz and R.~Willey, \emph{Monte carlo simulation calculation of the critical coupling constant for two-dimensional continuum $\varphi^4$ theory}, {\emph{Physical Review D} {\bfseries 58} (1998) 076003}.

\bibitem{stevenson1984gaussian}
P.~Stevenson, \emph{Gaussian effective potential: Quantum mechanics}, {\emph{Physical Review D} {\bfseries 30} (1984) 1712}.

\bibitem{stevenson1985gaussian}
P.~Stevenson, \emph{Gaussian effective potential. ii. $\lambda$ $\phi_4$ field theory}, {\emph{Physical Review D} {\bfseries 32} (1985) 1389}.

\bibitem{somma2005quantum}
R.D.~Somma, \emph{Quantum computation, complexity, and many-body physics}, {\emph{arXiv preprint quant-ph/0512209} (2005) }.

\bibitem{macridin2018digital}
A.~Macridin, P.~Spentzouris, J.~Amundson and R.~Harnik, \emph{Digital quantum computation of fermion-boson interacting systems}, {\emph{Physical Review A} {\bfseries 98} (2018) 042312}.

\bibitem{klco2019digitization}
N.~Klco and M.J.~Savage, \emph{Digitization of scalar fields for quantum computing}, {\emph{Physical Review A} {\bfseries 99} (2019) 052335}.

\bibitem{gleinig2021efficient}
N.~Gleinig and T.~Hoefler, \emph{An efficient algorithm for sparse quantum state preparation. in 2021 58th acm/ieee design automation conference (dac)},  2021.

\bibitem{trotter1959product}
H.F.~Trotter, \emph{On the product of semi-groups of operators}, {\emph{Proceedings of the American Mathematical Society} {\bfseries 10} (1959) 545}.

\bibitem{suzuki1976generalized}
M.~Suzuki, \emph{Generalized trotter's formula and systematic approximants of exponential operators and inner derivations with applications to many-body problems}, {\emph{Communications in Mathematical Physics} {\bfseries 51} (1976) 183}.

\end{thebibliography}\endgroup

\end{document}